\documentstyle[12pt,osa]{revtex}

\begin{document}

\title{THE ENERGY-MOMENTUM TENSOR FOR THE GRAVITATIONAL FIELD.}
\author{S.V. Babak and L.P. Grishchuk}
\address{Department of Physics and Astronomy, Cardiff University,\\
 PO Box 913, CF2 3YB, UK.\\
{\rm Email: babak@astro.cf.ac.uk ;  grishchuk@astro.cf.ac.uk
}}
\maketitle
\begin{abstract}
 The field theoretical description of the general relativity (GR) is further developed.
 The action for the gravitational field and its sources is given explicitely.
The equations of motion and the energy-momentum tensor for the gravitational field
are derived by applying the variational principle. We have succeeded in
constructing the unique gravitational energy-momentum tensor which is
1) symmetric, 2) conserved due to the field equations, and 3)~contains
 not higher than the first order derivatives of the field variables.
 It is shown that the Landau-Lifshitz pseudotensor is an object most closely
 related to the derived energy-momentum tensor. 
\end{abstract}

\section{Introduction}

 For any Lagrangian based field theory a 'metrical'
energy-momentum tensor is defined by

\begin{equation}
T^{\mu\nu} = -\frac 2{\sqrt{-g}} \frac{\delta L}{\delta g_{\mu\nu}}, \label{1}
\end{equation}
where $g_{\mu\nu}$ is the metric tensor, $L$ is the
Lagrangian density of the field and $g$ is the determinant of
$g_{\mu\nu}$.  
The field equations (the Euler-Lagrange equations) are derived by applying the
variational principle:

\begin{equation}
\frac {\delta L}{\delta \phi_A}=0,
\end{equation}
where $\phi_A$ are field variables. In the geometrical description
of GR,  one identifies the gravitational field with the geometry of the curved
space-time, and the metric tensor $g_{\mu\nu}$ plays the dual role.
On one side, the metric tensor $g_{\mu\nu}$ defines the  metrical relations
in the space-time, on the other side,
the components of the $g_{\mu\nu}$ are regarded as the gravitational field
variables.
The variation of $L$ with respect to $g_{\mu\nu}$
gives the Einstein equations.
If one defines the gravitational energy-momentum tensor according to (\ref{1}),
it becomes equal to zero due to the Einstein field equations.

The quantities which are normally being used are various
energy-momentum pseudotensors. They are known to be
unsatisfactory. The lack of a rigorously defined energy-momentum tensor
for the gravitational field becomes especially acute in problems such as
quantization of cosmological perturbations. We believe that the difficulty
lies in the way we treat the gravitational field, not in the nature of
gravity as such.

The Einstein gravity can be perfectly well formulated as a field theory
in the Minkowski space-time. This allows us to introduce a perfectly
acceptable energy-momentum tensor for the gravitational field.

\section{The field theoretical approach to gravity}

The field theoretical approach treats gravity as
a nonlinear tensor field $h^{\mu\nu}$ given in the Minkowski space-time.
This approach has a long and fruitful history (see for example
\cite{Feynman,Deser,GPP}).
However, the previous
work was focused on demonstrating that the Einstein equations are,
in a sense, most natural and unavoidable, rather than on any practical
applications of this approach.

We will follow a concrete representation developed in \cite{GPP}. The
gravitational field is denoted by $h^{\mu\nu}$, and the metric tensor
of the Minkowski space-time by $\gamma^{\mu\nu}$ .
The Christoffel symbols associated with $\gamma^{\mu\nu}$
are denoted by $C^{\tau}_{\ \mu\nu}$, the covariant derivatives
with respect to $\gamma^{\mu\nu}$ by ";" and the curvature tensor
of the Minkowski space-time is

\begin{equation}
\breve R_{\alpha\beta\mu\nu}(\gamma^{\rho\sigma})=0. \label{r}
\end{equation}

\subsection{The variational principle}

The action for the gravitational field is

\begin{equation}
S^g=- \frac1{2c\kappa} \int L^g \;d^4x, \nonumber
\end{equation}
where $\kappa = 8\pi G / c^4$.

The energy-momentum tensor for the gravitational field
is defined in the traditional manner, as the variational derivative with
respect to the metric tensor $\gamma_{\mu\nu}$:

\begin{equation}
\kappa t^{\mu\nu}|_v = -\frac 1{\sqrt{-\gamma}} \frac{\delta L^g}{\delta \gamma _{\mu\nu}},
\label{4}
\end{equation}
The field equations are derived by applying the variational principle 
with respect to the field variables $h^{\mu\nu}$ (gravitational potentials).
It is convenient (but not necessarily) to consider the generalised momenta
$P^{\alpha}_{\ \mu\nu}$, canonically conjugated 
to the generalised coordinates $h^{\mu\nu}$, as independent
variables . This is an element of the Hamiltonian formalism, which is known
also as the first order variational formalism.
 So, the field equations, in the framework of the first order formalism, are
\begin{equation}
\frac{\delta L^g}{\delta h^{\mu\nu}}= 0, \label{5}
\end{equation}
\begin{equation}
\frac{\delta L^g}{\delta P^{\tau}_{\ \mu\nu}} =0. \label{6}
\end{equation}
The  concrete Lagrangian density for the gravitational field used in
\cite{GPP} is given by: 
\begin{equation}
L^{g}= \sqrt{-\gamma} \left({h^{\rho\sigma}}_{;\alpha}P^{\alpha}_{\ \rho\sigma}
- \frac 1{2}{\Omega^{\rho\sigma\alpha\beta}}_{\omega\tau}P^{\tau}_{\ \rho\sigma}P^{\omega}_{\ \alpha\beta}
\right),\label{7}
\end{equation}
where
\begin{eqnarray}
{\Omega^{\rho\sigma\alpha\beta}}_{\omega\tau} &\equiv&
 \frac 1{2}
 [ (\gamma^{\rho\alpha}+h^{\rho\sigma}){Y^{\sigma\beta}}_{\omega\tau} +
(\gamma^{\sigma\alpha}+h^{\sigma\alpha}){Y^{\rho\beta}}_{\omega\tau}+
\nonumber\\ 
& &(\gamma^{\rho\beta}+ h^{\rho\beta}){Y^{\sigma\alpha}}_{\omega\tau}+
 (\gamma^{\rho\alpha} + h^{\rho\alpha}){Y^{\sigma\beta}}_{\omega\tau}]\label{o}
\end{eqnarray}
and
$$
{Y^{\rho\alpha}}_{\sigma\beta} \equiv
\delta^{\rho}_{\sigma}\delta^{\alpha}_{\beta} -\frac 1{3} \delta^{\rho}_{\beta}
\delta^{\alpha}_{\sigma}.
$$

\subsection{The field equations}

By direct calculation of the variational derivatives of the Lagrangian
density (\ref{7}) one can obtain the field equations:
\begin{equation}
\frac1{\sqrt{-\gamma}}\frac{\delta L^g}{\delta h^{\mu\nu}}\equiv
 - \left( P^{\alpha}_{\ \mu\nu ;\alpha} + P^{\alpha}_{\ \mu\beta}
P^{\beta}_{\ \nu\alpha} -\frac1{3} P_{\mu} P_{\nu}\right)= 0, \label{8}
\end{equation}
\begin{equation}
\frac1{\sqrt{-\gamma}}\frac{\delta L^g}{\delta P^{\tau}_{\ \mu\nu}}\equiv
 {h^{\mu\nu}}_{;\tau} - {
\Omega^{\mu\nu\alpha\beta}}_{\omega\tau}P^{\omega}_{\ \alpha\beta}=0,\label{9}
\end{equation}
where $P_{\rho}\equiv P^{\alpha}_{\ \rho\alpha}$.
Equation (\ref{9}) provides the link between $P^{\alpha}_{\ \mu\nu}$
and $h^{\mu\nu}$:

\begin{equation}
 {h^{\mu\nu}}_{;\tau} =
 {\Omega^{\alpha\beta\mu\nu}}_{\tau\omega}P^{\omega}_{\ \alpha\beta},\label{9b}
\end{equation}
One can also resolve  equation (\ref{9b}) in terms of $P^{\tau}_{\ \mu\nu}$:

\begin{equation}
P^{\tau}_{\ \mu\nu} = {\Omega^{-1}_{\rho\sigma\mu\nu}}^{\tau\omega}
{h^{\rho\sigma}}_{;\omega}, \label{p}
\end{equation}
where ${\Omega^{-1}_{\rho\sigma\mu\nu}}^{\tau\omega}$ is the inverse
matrix to the  matrix
${\Omega^{\alpha\beta\mu\nu}}_{\tau\omega}$, namely

\begin{equation}
{\Omega^{\mu\nu\alpha\beta}}_{\omega\tau}
{\Omega^{-1}_{\rho\sigma\mu\nu}}^{\tau\psi} \equiv \frac 1{2}
\delta^{\psi}_{\omega}(\delta^{\alpha}_{\rho}\delta^{\beta}_{\sigma}+
\delta^{\alpha}_{\sigma}\delta^{\beta}_{\rho}).\label{o-1} 
\end{equation}

The field equations (\ref{8})  could be also  derived
using the Lagrangian formalism, known also as a second
order  variational formalism.
To implement  this one has to consider $P^{\alpha}_{\ \mu\nu}$ as known
functions of $h^{\mu\nu}$ and $h^{\mu\nu}\ _{;\alpha}$ (see (\ref{p}))
and substitute $P^{\tau}_{\ \mu\nu}$ into the Lagrangian (\ref{7}). After
performing  this substitution, the Lagrangian  takes the following form
\begin{equation}
L^{g}=\frac 1 {2} \sqrt{-\gamma} {\Omega^{-1}_{\rho\sigma\alpha\beta}}^{\omega\tau}
{h^{\rho\sigma}}_{;\tau} {h^{\alpha\beta}}_{;\omega} ,
\label{3n}
\end{equation}
which is explicitely quadratic in term of "velocities"  ${h^{\mu\nu}}_{;\tau}$.
The field equations, in framework of the second order variational formalism,
are

\begin{equation}
\frac{\delta L^g}{\delta h^{\mu\nu}}= 0 \label{5n}
\end{equation}
These equations are exactly the same as equations (\ref{8}), if one takes into account the link
(\ref{p}).

\subsection{Connection to the geometrical GR}

Equations (\ref{8}), (\ref{9}) are fully equivalent to the Einstein equations in the
geometrical approach. To demonstrate the equivalence, one has to introduce
new quantities $g^{\mu\nu}$ according to the definition:

\begin{equation}
 \sqrt{-g}g^{\mu\nu}=\sqrt{-\gamma}(\gamma^{\mu\nu}+ h^{\mu\nu})\label{g}
\end{equation}
and interpret $g_{\mu\nu}$ as the metric tensor of a curved space-time.
The matrix $g_{\mu\nu}$ is the inverse matrix
to $g^{\mu\nu}$
\begin{equation}
g_{\mu\alpha}g^{\nu\alpha}=\delta^{\nu}_{\mu} \label{g-1}
\end{equation}
and  $g$ is the determinant of $g_{\mu\nu}$.
 
The Christoffel symbols $\Gamma^{\alpha}_{\ \mu\nu}$ constructed from
$g_{\mu\nu}$, with (\ref{p}) and (\ref{g}) taken into account, have the
following form

\begin{equation}
\Gamma^{\alpha}_{\ \mu\nu} = C^{\alpha}_{\ \mu\nu} 
 -P^{\alpha}_{\ \mu\nu} + \frac 1{3} \delta^{\alpha}_{\mu}
P_{\nu} + \frac 1{3} \delta^{\alpha}_{\nu} P_{\mu}. \label{G}
\end{equation}
 The vacuum Einstein's equations

\begin{equation}
R_{\mu\nu} = {\Gamma^{\alpha}_{\ \mu\nu}}_{,\alpha} -\frac1{2}
\Gamma_{\mu,\nu} -\frac1{2} \Gamma_{\nu,\mu}
+ \Gamma^{\alpha}_{\ \mu\nu}\Gamma_{\alpha}-
\Gamma^{\alpha}_{\ \mu\beta}\Gamma^{\beta}_{\ \nu\alpha}=0, \label{ein}
\end{equation}
with  $\Gamma^{\alpha}_{\ \mu\nu}$ taken from (\ref{G}), reduce to

\begin{equation}
R_{\mu\nu} = \breve R_{\mu\nu}- \left( P^{\alpha}_{\ \mu\nu ;\alpha} + P^{\alpha}_{\ \mu\beta}
P^{\beta}_{\ \nu\alpha} -\frac1{3} P_{\mu} P_{\nu}\right)= 0,
\end{equation}
which are exactly the field equations (\ref{8}), because $\breve R_{\mu\nu}=0$.

\section{The energy-momentum tensor for the gravitational field}

In general, the energy-momentum tensor derived from the Lagrangian (\ref{7})
according to the definition (\ref{4}) contains  the
  second order derivatives of the gravitational potentials $h^{\mu\nu}$:

\begin{eqnarray}
\kappa t^{\mu\nu}|_v &=&
\frac1{2} \gamma^{\mu\nu}{h^{\rho\sigma}}_{;\alpha}
P^{\alpha}_{\ \rho\sigma} + [\gamma^{\mu\rho}\gamma^{\nu\sigma} -
\frac1{2}\gamma^{\mu\nu}(\gamma^{\rho\sigma} + h^{\rho\sigma})]
(P^{\alpha}_{\ \rho\beta}P^{\beta}_{\ \sigma\alpha} -
 \nonumber \\
& & \frac1{3} P_{\rho}P_{\sigma}) + Q^{\mu\nu}, \label{tv}
\end{eqnarray}
where

\begin{equation}
Q^{\mu\nu}= \frac1{2}(\delta^{\mu}_{\rho}\delta^{\nu}_{\sigma}+
\delta^{\nu}_{\rho}\delta^{\mu}_{\sigma})
[-\gamma^{\rho\alpha}
h^{\beta\sigma}P^{\tau}_{\ \alpha\beta} +(\gamma^{\alpha\tau}h^{\beta\rho} -
\gamma^{\alpha\rho}h^{\beta\tau})P^{\sigma}_{\ \alpha\beta}]_{;\tau}.
\end{equation}

Some of the second order derivatives (but not all) can be excluded by using the
field equations. 
We regard an energy-momentum tensor physically satisfactory if it does not
depend on the second derivatives of the field potentials.
  The remaining freedom in  the  Lagrangian (\ref{7}), which preservs the field equations
(\ref{8}), (\ref{9}), allows us to build such an object.

Our aim is to construct a symmetric conserved energy-momentum tensor 
 which does not contain the higher than the first order derivatives of the
gravitational potentials $h^{\mu\nu}$. These are the demands which should be
met by a fully acceptable energy-momentum tensor. 

To derive such an energy-momentum tensor we use the most  general
Lagrangian density consistent with the Einstein equations (\ref{8}),
(\ref{9}) and with the variational procedure:

\begin{eqnarray}                             
L^{g} = \sqrt{-\gamma} \left[{h^{\rho\sigma}}_{;\alpha}P^{\alpha}_{\ \rho\sigma}
- \frac 1{2}{\Omega^{\rho\sigma\alpha\beta}}_{\omega\tau}P^{\tau}_{\ \rho\sigma}P^{\omega}_{\ \alpha\beta}\right] + 
 \sqrt{-\gamma}\Lambda^{\alpha\beta\rho\sigma} \breve R_{\alpha\rho\beta\sigma},\label{l'}
\end{eqnarray}
where $\breve R_{\alpha\rho\beta\sigma}$ is the curvature tensor (\ref{r})
constructed from $\gamma_{\mu\nu}$. We have added zero to the original Lagrangian,
but this is a typical  way of incorporating  a constraint (in our case,
$\breve R_{\alpha\rho\beta\sigma}=0$) by means of the undetermined
Lagrange multipliers.
The multipliers $\Lambda^{\alpha\mu\beta\nu}$ form a tensor which depends on $\gamma^{\mu\nu}$
and $h^{\mu\nu}$. The added term affects the energy-momentum tensor,
but does not change the field equations.

The field equations are derived from (\ref{l'}) using the variational principle
 as it was described above (see (\ref{5}), (\ref{6})). The equations derived
from the new Lagrangian  are exactly the same as original equations (\ref{8}),
(\ref{9}).

The energy-momentum tensor directly derived from (\ref{l'}) according to
(\ref{4}) is now modified as compared with (\ref{tv}):
\begin{eqnarray}
\kappa t^{\mu\nu}|_v &=&
\frac1{2} \gamma^{\mu\nu}{h^{\rho\sigma}}_{;\alpha}
P^{\alpha}_{\ \rho\sigma} + [\gamma^{\mu\rho}\gamma^{\nu\sigma} -
\frac1{2}\gamma^{\mu\nu}(\gamma^{\rho\sigma} + h^{\rho\sigma})]
(P^{\alpha}_{\ \rho\beta}P^{\beta}_{\ \sigma\alpha} -
 \nonumber \\
& & \frac1{3} P_{\rho}P_{\sigma}) + Q^{\mu\nu} +
\frac1{\sqrt{-\gamma}} \frac{\partial [\sqrt{-\gamma} \gamma_{\alpha\tau}
\Lambda^{\tau\beta\rho\sigma}]}{\partial \gamma_{\mu\nu}}
\breve R^{\alpha}_{\ \rho\beta\sigma} -
\nonumber\\
& & (\Lambda^{\nu\beta\mu\alpha} + \Lambda^{\mu\beta\nu\alpha})_{;\alpha ;\beta}.\label{tv'}
\end{eqnarray}
This tensor still contains the second derivatives
of the gravitational potentials $h^{\mu\nu}$, but the multipliers
$\Lambda^{\alpha\beta\rho\sigma}$
can be chosen in  such a way that
the remaining terms which 
contain the second derivatives of $h^{\mu\nu}$ (and which could not be
excluded by using the field equations) can now be removed. We have shown
that the unique choice of
$\Lambda^{\alpha\beta\rho\sigma}$ is
\begin{equation}
\Lambda^{\alpha\beta\rho\sigma}=
-\frac 1{4} (h^{\alpha\beta}h^{\rho\sigma} -h^{\alpha\sigma}h^{\beta\rho}).
\end{equation}

 As a result we have obtained the  energy-momentum tensor (which we 
call the true energy-momentum tensor), which is 
\begin{itemize}
\item    1) symmetric,

\item    2) conserved, ${t^{\mu\nu}}_{;\nu}=0$, as soon as the field equations
         are satisfied,

\item    3) does not contain any derivatives of $h^{\mu\nu}$ higher than the
         first order.

\end{itemize}
It is necessarily to emphasize that $t^{\mu\nu}$ is a tensor (not a pseudotensor),
so that it transforms as a tensor under arbitrary coordinate transformations.

%\newpage
The true energy-momentum tensor has the following final form:

\begin{eqnarray}
\kappa t^{\mu\nu} &=&
\frac1{4} [2 {h^{\mu\nu}}_{;\rho} 
{h^{\rho\sigma}}_{;\sigma} 
-2{h^{\mu\alpha}}_{;\alpha}{h^{\nu\beta}}_{;\beta}
+
2g^{\rho\sigma}g_{\alpha\beta} {h^{\nu\beta}}_{;\sigma} 
{h^{\mu\alpha}}_{;\rho} + \nonumber \\ 
& & g^{\mu\nu}g_{\alpha\rho} {h^{\alpha\beta}}_{;\sigma} 
{h^{\rho\sigma}}_{;\beta}-
  2g^{\mu\alpha}g_{\beta\rho} {h^{\nu\beta}}_{;\sigma}{h^{\rho\sigma}}_
{;\alpha} - \nonumber \\
& & 2g^{\nu\alpha} g_{\beta\rho} {h^{\mu\beta}}_{;\sigma}
{h^{\rho\sigma}}_{;\alpha} + \frac1{4}(2g^{\mu\delta}g^{\nu\omega} -g^{\mu\nu}g^{\omega\delta})
(2g_{\rho\alpha}g_{\sigma\beta} - \nonumber \\
& & g_{\alpha\beta}g_{\rho\sigma})
{h^{\rho\sigma}}_{;\delta}
{h^{\alpha\beta}}_{;\omega}],\label{49}
\end{eqnarray}
where $g_{\alpha\beta}$ and $g^{\alpha\beta}$ are defined by (\ref{g}) and
(\ref{g-1}).

It is important to point out that $t^{\mu\nu}$ can be numerically related
with the Landau-Lifshitz pseudotensor \cite{LL}. If one introduces the
quantities $g^{\mu\nu}$ according to (\ref{g}) and uses the Lorentzian
coordinate frame of reference
($\gamma_{00}=1, \gamma_{11}=\gamma_{22}=\gamma_{33}=-1$)
 one obtains the relationship between $t^{\mu\nu}_{LL}$ (the Landau-Lifshitz
 pseudotensor) and $t^{\mu\nu}$:

\begin{equation}
t^{\mu\nu} = (-g) t^{\mu\nu}_{LL}.
\end{equation}

\section{Gravitational field with matter sources}

So far we were working with gravitational equations without matter sources,
but all procedure cosidered above is also true for gravitational field in
presence of matter. In order to maintain equivalence with the Einstein equations
in its geometrical form one needs to ensure the universal coupling of gravity
to other physical fields (matter sources). The universal coupling between
gravity and matter is presented as the following functional dependence of
the matter Lagrangian density:

\begin{equation}
L^m=L^m(\sqrt{-\gamma} (\gamma^{\mu\nu} + h^{\mu\nu}); \phi_A;\phi_{A,\alpha}),
\label{57}
\end{equation}
where $\phi_A$ is an arbitrary matter field.
Then, the energy-momentum tensor $\tau^{\mu\nu}$ of matter sources and their interaction
with gravity (derived from $L^m$ according
to the same rule (\ref{4})) participates on equal footing with $t^{\mu\nu}$
as the right hand side of the field equations:

\begin{eqnarray}
[(\gamma^{\mu\nu}+ h^{\mu\nu})(\gamma^{\alpha\beta}+ h^{\alpha\beta})-
(\gamma^{\mu\alpha}+h^{\mu\alpha})
(\gamma^{\nu\beta}+h^{\nu\beta})]_{;\alpha;\beta} = \nonumber \\
=\frac{16 \pi G}{c^4}[t^{\mu\nu} + \tau^{\mu\nu}]. \label{14}
 \end{eqnarray}
These equations are totally equivalent to the geometrical Einstein's
equations

\begin{equation}
R^{\mu\nu}-\frac 1{2} g^{\mu\nu}R = \kappa T^{\mu\nu},
\end{equation}
where $g^{\mu\nu}$ is defined by (\ref{g}), $R^{\mu\nu}$ is the Ricci
tensor constructed from $g^{\mu\nu}$, and $T^{\mu\nu}$ is derived from
the same matter Lagrangian (\ref{57}) according to the rule (\ref{1}).

\section{Conclusion}

The derived energy-momentum tensors $t^{\mu\nu}$ and $\tau^{\mu\nu}$
are fully satisfactory from the physical point of view and should be useful
tools in practical applications.

\section*{Acknowledgments}
S. Babak was partially supported by ORS Award grant $N$ 97047008.

\end{document}